\documentclass[12pt]{article}
\usepackage{amssymb}
\begin{document}

\begin{center}
{\bf \Large Tests of conjectures on multiple Watson values}\\[10pt]
{\bf \large David Broadhurst}\\[5pt]
{Department of Physical Sciences, Open University, Milton Keynes MK7 6AA, UK\\
Institut f\"{u}r Mathematik und Institut f\"{u}r Physik, Humboldt-Universit\"{a}t zu Berlin}\\[5pt]
29 April 2015
\end{center}

{\large
I define multiple Watson values (MWVs) as iterated integrals,
on the interval $x\in[0,1]$, of the 6 differential forms
$A=d\log(x)$, $B=-d\log(1-x)$,
$T=-d\log(1-z_1x)$,
$U=-d\log(1-z_2x)$,
$V=-d\log(1-z_3x)$ and
$W=-d\log(1-z_4x)$,
where $z_1=\gamma^2$, $z_2=\gamma/(1+\gamma)$,
$z_3=\gamma^2/(1-\gamma)$ and $z_4=\gamma=2\sin(\pi/14)$ solves the
cubic $(1-\gamma^2)(1-\gamma)=\gamma$.
Following a suggestion by Pierre Deligne, I conjecture that
the dimension
of the space of ${\mathbb Z}$-linearly independent
MWVs of weight $w$ is the number $D_w$ generated by
$1/(1-2x-x^2-x^3)=1+\sum_{w>0}D_w x^w$.
This agrees with 6639 integer relation searches, of dimensions up
to $D_5+1=85$, performed
at 2000-digit precision, for $w<6$.

\newpage
\section{Introduction}

In 1937, G.N.~Watson, then
Mason Professor of Mathematics
at the University of Birmingham~\cite{JMW} in the English midlands,
published (with odd spelling) 
{\em A note on Spence's logarithmic transcendant}~\cite{W37}
wherein he considered the dilogarithm ${\rm Li}_2(y)=\sum_{n>0}y^n/n^2$
at 6 points in the algebraic number field defined by the cubic $C(x)=(1-x^2)(1-x)-x$, namely
at $y\in\{\alpha,\beta,\gamma,\alpha^2,\beta^2,\gamma^2\}$ with
$\alpha=1-\gamma^2$, $\beta=1-\gamma$ and $\gamma=2\sin(\pi/14)$
solving $C(\gamma)=0$.
In particular, he proved that the three combinations
${\rm Li}_2(\alpha)-{\rm Li}_2(\alpha^2)$,
$2\,{\rm Li}_2(\beta)+{\rm Li}_2(\beta^2)$ and
$2\,{\rm Li}_2(\gamma)+{\rm Li}_2(\gamma^2)$
evaluate  to combinations of $\pi^2$ and products
of the logarithms of $\gamma$ and $1-\gamma$.
In~\cite{Lewin82}, Leonard Lewin remarked that
``Watson indicated that he had long suspected the existence of a certain result,
and although his eventual proof is easy enough to follow,
it is clear that it was not all that easy to come by.''

My interest in Watson's paper was rekindled by a recent suggestion~\cite{letter2}
by Pierre Deligne that for prime $p=2n+3$ there is a set of
words in an alphabet of $p-1$ letters such that iterated integrals
on the interval $x\in[0,1]$ encoded by words of length $w$
are ${\mathbb Q}$-linear combinations of $D_w$ basis
terms, where $D_w$ is the coefficient of $x^w$ in the
expansion of $1/(1-n x -x^2-x^3)$.

For $p=3$, we have
the alphabet $\{A,B\}$ of multiple zeta values~\cite{BBV} with
$A=d\log(x)$ and $B=-d\log(1-x)$.  For these, the
dimensions are enumerated by the Padovan numbers,
obtained by expanding $1/(1-x^2-x^3)$.

For $p=5$, we have
the alphabet $\{A,B,F,G\}$ of multiple  Landen values~\cite{MLV} with
$F=-d\log(1-\rho^2x)$ and $G=-d\log(1-\rho x)$, where
$\rho=2\sin(\pi/10)=(\sqrt{5}-1)/2$ is the golden section.
For these, I conjectured that the dimensions are enumerated by the
tribonacci numbers, obtained by expanding $1/(1-x-x^2-x^3)$,
and gave evidence that this is the case for $w<9$.

This paper concerns $p=7$, with an alphabet $\{A,B,T,U,V,W\}$, where
\begin{eqnarray}
T&=&-d\log(1-z_1x),\quad{\rm with}\quad z_1=\gamma^2,\label{z1}\\
U&=&-d\log(1-z_2x),\quad{\rm with}\quad z_2=\gamma/(1+\gamma),\label{z2}\\
V&=&-d\log(1-z_3x),\quad{\rm with}\quad z_3=\gamma^2/(1-\gamma),\label{z3}\\
W&=&-d\log(1-z_4x),\quad{\rm with}\quad z_4=\gamma=2\sin(\pi/14),\label{z4}
\end{eqnarray}
are carefully chosen to have $0<z_k<\frac12$. Section~2 gives a theorem and two conjectures.
Section~3 presents evidence and Section~4 offers comments.

\newpage

\section{Theorem and conjectures}

{\bf Theorem 1:} Let $z_{k+4}=1-z_k$, for $k=1$ to 4. Then  7 linearly independent rational combinations
of ${\rm Li}_2(z_k)$, with $k=1$ to 8, evaluate to combinations of $\pi^2$ and products of logarithms.

{\bf Proof:} Watson considered ${\rm Li}_2(y)$ with
$y\in\{\alpha,\beta,\gamma,\alpha^2,\beta^2,\gamma^2\}$. From these 6 arguments
we obtain 3 more, with $y\in\{\alpha/(1+\alpha),\beta/(1+\beta),\gamma/(1+\gamma)\}$,
using $2\,{\rm Li}_2(y)\simeq{\rm Li}_2(y^2)+2\,{\rm Li}_2(y/(1+y))$, where $\simeq$
denotes neglect of $\pi^2$ and product of logs. Then ${\rm Li}_2(y)+{\rm Li}_2(1-y)\simeq0$
gives a further 3, with  $y\in\{1/(1+\alpha),1/(1+\beta),1/(1+\gamma)\}$. Simple algebra
shows that only 8 of the 12 arguments are distinct and that these are the
values of $z_k$ for $k=1$ to 8. Watson proved the 3 relations
${\rm Li}_2(\alpha)\simeq{\rm Li}_2(\alpha^2)$,
$2\,{\rm Li}_2(\beta)+{\rm Li}_2(\beta^2)\simeq0$ and
$2\,{\rm Li}_2(\gamma)+{\rm Li}_2(\gamma^2)\simeq0$. Then 4 coincidences
of 12 arguments increase the number of relations to 7. Concretely, we have
${\rm Li}_2(z_k)\simeq c_k{\rm Li}_2(\gamma)$ with integers,
{\tt -2, 2, -2, 1, 2, -2, 2, -1,} giving $c_k$, for $k=1$ to 8.~$\blacksquare$

Multiple Watson values (MWVs) are iterated integrals, on $x\in[0,1]$, defined
by words in the alphabet $\{A,B,T,U,V,W\}$ that neither begin with $B$ nor end in $A$.
The weight $w$ of a MWV is the number of letters in the word and its depth $d$
is the number not equal to $A$. Hence $\zeta(2)=Z(AB)$ and ${\rm Li}_2(\gamma)=Z(AW)$
are MWVs with $w=2$ and $d=1$. In general, a MWV of weight $w$
and depth $d$ is a $d$-fold nested sum
of the form
\begin{equation}
{\rm Li}_{a_1,a_2,\ldots,a_d}(y_1,y_2,\ldots,y_d)\equiv\sum_{n_1>n_2>\ldots>n_d>0}
\quad\prod_{j=1}^d\frac{y_j^{n_j}}{n_j^{a_j}}
\label{Li}
\end{equation}
with $w=\sum_j a_j$ and arguments $y_j$ in Watson's real subfield of
the cyclotomic field of 7th roots of unity,
defined by the cubic $(1-x^2)(1-x)-x$.
Thus for example $Z(AVWAAB)={\rm Li}_{2,1,3}(z_3,z_4/z_3,1/z_4)$.
MWVs enjoy a shuffle algebra, as iterated integrals, but the stuffle~\cite{trans} algebra
of nested sums does not close. For example, the shuffle product
$Z(T)Z(AU)=Z(TAU)+Z(ATU)+Z(AUT)$ gives a sum of MWVs, but the depth-1 term
in the corresponding
stuffle product ${\rm Li}_1(z_1){\rm Li}_2(z_2)
={\rm Li}_{1,2}(z_1,z_2)+{\rm Li}_{2,1}(z_2,z_1)+{\rm Li}_3(z_1z_2)$ is not a MWV.
This makes it quite difficult to prove that all 25 MWVs of weight 2 are ${\mathbb Q}$-linear
combinations of the basis $\{Z(AB),\,Z(AT),\,Z(TT),\,Z(T)Z(U),\,Z(UU)\}$, which Theorem~1
shows to be sufficient for the reduction of the depth-1 words at $w=2$. Thus I relied
on empirical methods, using the {\tt lindep} procedure of {\tt Pari-GP}~\cite{GP},
to test the following conjectures.

{\bf Conjecture 1:} The dimension
of the space of ${\mathbb Z}$-linearly independent
MWVs of weight $w$ in the alphabet $\{A,B,T,U,V,W\}$
is the number $D_w$ generated by
$1/(1-2x-x^2-x^3)=1+\sum_{w>0}D_w x^w$.

There is little hope of proving this in the foreseeable future.
Yet a proof that $D_w$ is an upper bound for the dimension
may be within reach~\cite{letter2}.

My choice of alphabet $\{A,B,T,U,V,W\}$  was made after
receiving~\cite{letter2} from Deligne, who was at first rather
reluctant to accept this construction as conforming to his ideas.
There was a good reason for my systematic choice, which selects
those 4 of Watson's 8 points that have $z_k<\frac12$,
as in ~(\ref{z1},\ref{z2},\ref{z3},\ref{z4}). This
makes numerical computation of MWVs far easier than for words that
involve letters with $1>z_k>\frac12$. Let's call the latter $t,u,v,w$,
obtained from $z_{k+4}=1-z_k$, for $k=1$ to 4.

 {\bf Conjecture 2:}
 Finite iterated integrals in the alphabets $\{A,B,t,u,v,w\}$
and $\{A,B,U,u,W,w\}$ are ${\mathbb Q}$-linear combinations of MWVs.
Moreover,  the dimensions for those alphabets are the same as in Conjecture~1.

Deligne has told me that he prefers  $\{A,B,U,u,W,w\}$ to the
computationally more convenient alphabet  of Conjecture~1.

\section{Evidence}

Being, by upbringing, an empiricist, I wished to test the conjectures
for the largest weight that is feasible. Since $D_6=214$ is infeasibly
large for {\tt lindep},  that meant restricting
attention to $w<6$. The tests at $w=5$, with $D_5=84$, are already more
demanding than those that I performed for multiple Deligne values~\cite{MDV},
at $w=11$, and multiple Landen values~\cite{MLV}, at $w=8$, where
the conjectured dimensions were 72 and 81, respectively.

A preliminary skirmish with a naive Hoffman-type basis,
based on words in a three-letter sub-alphabet,
led to large denominator-primes that caused me to miss a relation
at $w=4$, at 500-digit precision. That led to a false alarm
that instead of the predicted dimension $D_4=33$ the answer
might be the 9th Fibonacci number, $F_9=34$.

Fibonacci numbers with odd indices
are generated by $1/(1-x-x/(1-x))$, which gives
the sequence {\tt 2, 5, 13, 34, 89, 233,} for $w=1$ to 6,
while Conjecture~1 gives {\tt 2, 5, 13, 33, 84, 214},
and hence more relations for $w>3$. Running
at 2000-digit precision I found agreement with $D_4=33$.

Chastened by experience at $w=4$,  I resorted to
a method that seemed more likely to avoid large denominator
primes. As in~\cite{MLV}, I resolved not to restrict the primitives
to a sub-alphabet. Instead my Aufbau was based on ordering
primitive MWVs first by weight, $w$,  then by depth, $d$,
and finally, for each $w$ and $d$, by lexicographic order.
By conjecture, the number $N_w$ of primitives is
generated by
\begin{equation}
\prod_{w>0} (1-x^w)^{N_w}=1-2x-x^2-x^3
\label{prim}
\end{equation}
which, for $w=1$ to 6, gives the sequence {\tt 2, 2, 5, 9, 21, 42}.

A systematic
choice of primitive words is given by
$$\{T,\,U\},\quad\{AB,\,AT\},\quad\{A^2B,\,A^2T,\,A^2U,\,A^2V,\,A^2W\},$$
for $w=1$ to 3, respectively. Then at $w=4$, the primitives
$$\{A^3T,\,A^3U,\,A^3V,\,A^3W,\,A^2TB,\,A^2TU,\,A^2TV,\,A^2UB,\,A^2UT\}$$
solved the problem of large denominator primes.
Products of these 18 primitives supply 63 elements
of an 84-dimensional basis at $w=5$, leaving 21 new primitives 
to be determined.
Then 2000-digit precision was sufficient to find primitives $\{A^3Y_k\}$
with 21 two-letter endings, $Y_k$, in
\begin{eqnarray*}
&&\{
AB,\,
AT,\,
AU,\,
AV,\,
AW,\,
BT,\,
BU,\,
BV,\,
BW,\,
TB,\,
TU,\,
TV,\nonumber\\&&
\quad\quad\quad
UB,\,
UT,\,
UV,\,
UW,\,
VB,\,
VT,\,
VU,\,
VW,\,
WB\}
\end{eqnarray*}
which gives one denominator prime greater than 11, namely  9528587.

I then reduced all 1079 finite MWVs  with $w<5$ to bases of the conjectured sizes.
To test Conjecture~1 at $w=5$, it is sufficient to restrict attention
to 5-letter Lyndon words in $\{A,B,T,U,V,W\}$. If these reduce
to an 84-dimensional basis, then so do all finite weight-5 words,
since they are given by the shuffle algebra in terms of
Lyndon words and products. There are 1134 finite Lyndon words
with 5 letters taken from the 6-letter alphabet. Running {\tt lindep}
at 2000-digit precision, I found reductions for all of these
and hence successfully completed the tests of Conjecture~1, for $w<6$.

To test Conjecture 2, I reran exactly the same procedures for the
alphabets $\{A,B,t,u,v,w\}$ and $\{A,B,U,u,W,w\}$, obtaining, in each case,
1079 reductions to the $\{A,B,T,U,V,W\}$ bases with $w<5$
and 1134 reductions of weight-5 Lyndon words.

\section{Comments}

In~\cite{MLV}, I discovered a simple enumeration of multiple polylogarithms
in Landen's real subfield of the cyclotomic field of 5th roots of unity.
Credit for the generating function for $p=7$, in Conjecture~1,
goes to Deligne~\cite{letter2}, though he seemed, at first, to disapprove of my
systematically chosen alphabet, for the Watson problem.
The following seem to be pertinent points.
\begin{enumerate}
\item Quantum field theory has not yet produced periods that are multiple
polylogarithms at 7th roots of unity~\cite{GBU}. Instead the prime 7 appears
as an obstacle~\cite{BS} to polylogarithmic evaluation of Feynman
diagrams, signally, via a modular form of weight 3, the need for higher genus.
\item Deligne suggests~\cite{letter2} that for prime $p=2n+3$ there is
a $p-1$ letter alphabet with dimensions generated by $1/(1-n x-x^2-x^3)$. This soon leads to very
large dimensions for $p>7$.
\item The alphabet of Conjecture~1 is more accessible to computation than
those in Conjecture~2, which yield honorary MWVs, for $w<6$. For each of the
three alphabets, $1079+1134=2213$ integer relations attest to the fidelity of
the conjectures.
\end{enumerate}

{\bf Acknowledgment:} I am grateful to Pierre Deligne for a
lively and friendly exchange, in response to my discoveries
at $p=5$. His formidable powers of pure
thought~\cite{letter2} complement my interests in numerical
evidence~\cite{MDV,GBU,MLV,BS} and historical data~\cite{L6,L,W37}.

\raggedright

}
\begin{thebibliography}{99}

\bibitem{BBV}
J.~Bl\"umlein, D.J.~Broadhurst, J.A.M.~Vermaseren,
{\em The multiple zeta value data mine},
Comput.~Phys.~Commun., {\bf 181} (2010) 582--625,
{\tt arXiv:0907.2557}~.

\bibitem{trans}
J.M.~Borwein, D.M.~Bradley, D.J.~Broadhurst, P.~Lisonek,
{\em Special values of multiple polylogarithms},
Trans.~Amer.~Math.~Soc., {\bf 353} (2001) 907--941,
{\tt arXiv:math/9910045}~.

\bibitem{MDV}
D.~Broadhurst,
{\em Multiple Deligne values: a data mine with empirically tamed denominators},
{\tt arXiv:1409.7204}~.

\bibitem{GBU}
D.~Broadhurst,
{\em Polylogs of roots of unity: the good, the bad and the ugly},
talk at meeting on {\em Mathematical physics,  number theory and non-commutative geometry},
Vienna, 12 March 2015,
{\tt http://www.noncommutativegeometry.nl/esi2015/slides/}~.

\bibitem{MLV}
D.~Broadhurst,
{\em Multiple Landen values and the tribonacci numbers},
{\tt arXiv:1504.05303}~.

\bibitem{BS}
D.~Broadhurst, O.~Schnetz,
{\em Algebraic geometry informs perturbative quantum field theory},
Proc.~Sci., {\bf 211} (2014) 078,
{\tt arXiv:1409.5570}~.

\bibitem{letter2}
P~Deligne, letter to the author on 23 April 2015, in response to~\cite{MLV}.

\bibitem{L6}
J.~Landen,
{\em A new method of computing the sums of certain series},
Phil.~Trans.~Roy.~Soc., {\bf 51} (1759) 553--565.

\bibitem{L}
J.~Landen,
{\em Mathematical memoirs respecting a variety of subjects},
Volume 1, Nourse, London, 1780.

\bibitem{Lewin82}
L.~Lewin,
{\em The dilogarithm in algebraic fields},
J.~Austral.~Math.~Soc., {\bf A33} (1982) 302--330.

\bibitem{GP}
PARI~Group, PARI/GP version {\tt 2.5.0}, Bordeaux, 2011,
{\tt http://pari.math.u-bordeaux.fr/}~.

\bibitem{W37}
G.N.~Watson,
{\em A note on Spence's logarithmic transcendant},
Quart.~J.~Math., Oxford, {\bf 8} (1937) 39--42.

\bibitem{JMW}
J.M.~Whittaker,
{\em George Neville Watson, 1886-1965},
Biographical Memoirs of Fellows of the Royal Society,
{\bf 12} (1966) 520--530,
{\tt http://www.jstor.org/stable/769548}~.

\end{thebibliography}
\end{document}